\documentstyle[12pt]{article}

\begin{document}

%{\sf 

\begin{center}
\noindent
{\Large \bf Distributional Torsion of Charged Domain Walls With Spin
Density}\\[3mm]

by \\[0.3cm]

{\sl L.C. Garcia de Andrade}\\ 

\vspace{0.5cm}
Departamento de F\'{\i}sica
Te\'orica -- IF -- UERJ\\[-3mm] 
Rua S\~ao Francisco Xavier, 524\\[-3mm]
Cep 20550-003, Maracan\~a, Rio de Janeiro, RJ, Brasil\\[-3mm]
Electronic mail address: garcia@symbcomp.uerj.br\\[-3mm]

\vspace{2cm}
{\bf Abstract}
\end{center}
\paragraph*{}

An exact solution of Einstein-Cartan-Maxwell (ECM) field equations 
representing a charged domain wall given by the jump on the electric
charge and spin density across the wall is obtained from the Riemannian
theory of distributions. The Gauss-Coddazzi equations are used to
show that spin, charge and Cartan torsion increases the repulsive
character of the domain wall. Taub and Vilenkin walls are discussed as
well as their relations to wormhole geometry. The electric and
torsion fields are constants at the wall.
\vspace{0.5cm}
\noindent
{\bf PACS numbers:} \hfill\parbox[t]{13.5cm}{0420, 0450} 

\newpage
\section{Introduction}

\paragraph*{}
The motivation for the investigation of the domain
walls$^{\cite{um}}$ evolution stems for inflationary
universes$^{\cite{dois}}$. The physical motivation to investigate
charged domain walls$^{\cite{tres}}$ and domain walls with spinning
particles$^{\cite{quatro}\cite{cinco}}$ were provided mainly through
the works of C.A. Lopez$^{\cite{tres,cinco}}$ and \O . Gr\o
n$^{\cite{quatro}}$ in their study of repulsive gravitational
fields$^{\cite{seis,sete}}$. 
Their physical motivation was basic to build electron models
extending on Lorentz stress to General Relativity (GR). 
Since Hehl et al$^{\cite{oito}}$ have shown that Cartan torsion is
the geometrical interpretation of spin of the particles this provides a
natural motivation for the study of spinning particles in charged
domain walls. Therefore here I shall be concerned with an exact
static solution of ECM-field equations representing a charged planar
domain wall endowed with particles with spin. The Einstein 
cosmological constant although it is one of the responsibles for the
repulsive gravitation$^{\cite{nove}\cite{sete}}$ is absent here. In
fact the spin energy density plays the role of a ``cosmological
constant'' on a sort of torsion vacuum.

Here we use the fact discovered recently by myself and
Lemos$^{\cite{dez}}$ that in EC-gravity space-time defects can be
dealt with simply by substitution of the energy density and pressures
by effective quantities with the contribution of spin energy densities.
This allow us to reach a series of interesting physical conclusions
concerning the gravitationally repulsive character$^{\cite{onze}}$ of
domain walls as well as the horizon position in relation to
the planar wall. This paper in a certain extent completes and
generalizes my previous works on torsion walls$^{\cite{doze,treze}}$.

In section II we review ECM theory and apply in
the case of static Taub$^{\cite{quinze}}$ wall. It is important to
note that Taub wall in the context of General Relativity (GR) is the
well known no wall since $(\sigma =p=0)$ where $\sigma$ and $p$ are
respectively the energy density and the longitudinal pressure. This
situation is not trivial in EC-gravity since $\sigma^{eff}\equiv
(\sigma -2\pi S^2)$ and $p^{eff}\equiv p-2\pi S^2$
vanishing do not imply that $\sigma =0=p$. This is an important
different between our paper and the others dealing with domain walls 
in (GR). In section III some consequences of Section II as the
investigation of repulsive gravitation in
(EC)-gravity are found.

\section{Einstein-Cartan-Maxwell Gravitational Fields and
Electrostatic Domain Walls}

\paragraph*{}
Let us consider the formulation of ECM-field equations as given in
Tiwari and Ray$^{\cite{quinze}}$. The main difference is that here we
use (TR)-equations with plane
symmetry while in the TR-paper their solution is spherically
symmetric. In general the ECM-equations in the quasi-Einsteinian form
maybe written as 
\begin{equation}
G^b_a(\{\})=-8\pi\!\!\stackrel{eff}{T}{}^{\!\!b}_{\!\!a}
\end{equation}
which corresponds to
\begin{equation}
G^0_0(\{ \})=-2\mu_{zz}-3\mu^2_z+2\mu_z\nu_z=8\pi
[\sigma+\sigma_E-2\pi S^2]\delta (z)e^{2\nu}\equiv 8\pi
\sigma^{eff}\delta (z)
\end{equation}
\begin{equation}
G^2_2(\{ \})=G^3_3(\{ \})=-\mu_{zz}-\mu^2_{zz}-\nu_{zz}=-
8\pi [\sigma_E-2\pi S^2]e^{2\nu}\delta (z)=-8\pi p^{eff}_{||}e^{2\nu}\delta (z)
\end{equation}
\begin{equation}
G^1_1(\{ \})=-(\mu_z\nu_z+\mu^2_z)=8\pi[p^{eff}_{||}]\delta (z)e^{2\nu}=
8\pi [p_{\bot}](\sigma_E+2\pi S^3)]
\end{equation}

Here $\{ \}$ represents the Riemannian connection $\delta (z)$ is
the Dirac $\delta$-distribution and where we have taken the
Taub$^{\cite{dezesseis}}$ plane symmetric space-time as
\begin{equation}
ds^2=e^{2\nu }(-dt^2+dz^2)+e^{2\mu}(dx^2+dy^2)
\end{equation}
and we have also considered $S^0_{23}\equiv u^0S_{23}$ as the only
nonvanishing component of the spin density tensor and $u^0$ is the
zero-velocity of the four-velocity. Here $S_{23}$ is the spin angular
momentum tensor. In (GR) the orthogonal pressure $p_{\bot}$ vanishes.
But here what vanishes is the effective orthogonal pressure in (4)
\begin{equation}
p_{\bot}^{eff}\equiv 0\rightarrow p_{\bot}=(\sigma_E+2\pi S^2)
\end{equation}
where here $\sigma_E\equiv E^2/8\pi$ is the electrostatic surface
energy density. Since torsion does not couple with the electrostatic
field$^{\cite{dezesete}}$ the Maxwell equations  are the same as in
(GR). In our case we take the electric and torsion fields as
constants at the charged domain wall. Thus a simply first integration
of the system of (ECM)-differential system reads 
\begin{equation}
\mu^+_z=-\frac{1}{2}\ \nu^+_z
\end{equation}
\begin{equation}
-4\mu^+_z= 8\pi^{eff}e^{2\nu}
\end{equation}
\begin{equation}
2\mu^+_z-2\nu^+_z=-8\pi p^{eff}e^{2\nu}
\end{equation}

Now the Taub$^{\cite{dezesseis}}$ wal is defined in EC-gravity as 
\begin{equation}
\sigma^{eff}=p_{||}^{eff}=0
\end{equation}
which from (6) yields 
\begin{equation}
\sigma_E=+2\pi S^2
\end{equation}
or $E^2=16\pi^2S^2$ which implies
\begin{equation}
E=\pm 4\pi S
\end{equation}

Where the plus sign denotes the upper half Minkowski space-time. The
non-Riemannian space-time deffect here is obtained by gluing together
two half Minkowski spaces across spin-torsion-charge function.

Where the result (11) comes  
from classical electrostatics$^{\cite{dezoito}}$ is the electrostatic
field at the plane.

Therefore the spin density squared $S\equiv \sqrt{S^2}$ and $E=\pm 4\pi
\stackrel{0}{\sigma}_E$. Of course the plus and minus signs corresponds
to the upper and lower sides of the plane.

The Taub solution is the same as in (GR) and it
is$^{\cite{vinte}\cite{vum}}$ 
\begin{equation}
ds^2=-\frac{dt^2+dz^2}{\sqrt{1+Kz}}+(1+Kz)(dx^2+dy^2)
\end{equation}

Let us now solve eqns. (7-9) for the Vilenkin$^{\cite{um}}$ wall
where $\sigma^{eff}=-4p_{||}^{eff}$ which from (2) and (3) reads 
\begin{equation}
\sigma +\sigma_E-2\pi S^2=-4(\sigma_E-2\pi S^2)
\end{equation}
or
\begin{equation}
\sigma =-5(\sigma_E+2\pi S^2)
\end{equation}
for the surface energy density of the wall. By solving the eqns.
(7-9) one obtains
\begin{equation}
e^{2\nu}=-e^{-\mu}=(1-4\pi \sigma^{eff}|z|)^{-1/2}
\end{equation}
by substitution of the RHS of (2) into (16) one obtains the final
``Vilenkin'' charged domain wall with spin and torsion 
\begin{equation}
e^{2\nu}=e^{-\mu}\left(1-[4\pi \sigma +E^2+8\pi S^2]|z|\right)
\end{equation}
where $E^+=4\pi \stackrel{0}{\sigma}_E$ is the electrostatic field on
upper half of the planar wall. Notice that the gravitational
potential $g_{00}$ in (17) is written in terms of the electrostatic
potential when the weak field limit is taken. We may write the whole
energy density as $\sigma^{eff}=\sigma +2\sigma_E+2\pi S^2$.

\section{Gravitational Repulsion of the Charged Domain Wall With Spin and
Torsion}

\paragraph*{}
In this Section I shall review the Riemannian Gauss-Coddazzi
equations given in Ipser and Skivie$^{\cite{onze}}$. Since the only
modification of charged spinning particles domain wall is the
introduction of spin and electric surface charge densities into the
effective energy densities, the modification of physical conclusions
of (ECM)-domain walls is basically due to the changes in the
densities. Before the examination of the Gauss-Codazzi eqns. let us
consider the first important physical change due to the introduction
of spin, torsion and the electrostatic field into the domain wall.
The first change is obvious from the metric (17) and is given by the
``new horizon'' singularity 
\begin{equation}
g_{00}=\left(1-\left[4\pi \sigma
+E^2+8\pi^2S^2\right]|z|\right)^{-1/2}\rightarrow \infty 
\end{equation}
where
\begin{equation}
\sigma^{eff}=\sigma +2\sigma_E+2\pi S^2
\end{equation}
yielding 
\begin{equation}
|z|=\frac{1}{4\pi [\sigma^{eff}]}
\end{equation}

>From (19) one sees that $2\pi S^2$ term gives an extra spin
contribution to the wall surface density.

Expression (20) simply tell us that the position of the singularity
becomes closer to the wall due to the influences of charge and spin.

A similar situation appears in the case of the cosmological constant
in the case of plane symmetric exact solution of Einstein equations.

Since the first term on the RHS of (21) is positive if $E>0$, $\sigma
<0$ which reminds us of the wormwhole geometry$^{\cite{dezenove}}$.
Let us now turn to the Gauss-Coddazzi equations
\begin{equation}
^3R+\pi_{ij}\pi^{ij}-\pi^2=-2G_{ij}\xi^i\xi^j
\end{equation}
\begin{equation}
h_{ij}D_K\pi^{jK}-D_i\pi =G_{jK}h^j_i\xi^K
\end{equation}
where $^3R$ is the Ricci scalar curvature of the 3-dimensional
geometry $h_{ij}$ of the surface $S,$ $\pi \equiv \pi^i_i$ where.
\begin{equation}
\pi_{ij}\equiv D_i\xi_j=\pi_{ji}
\end{equation}
is the extrinsic curvature and 
\begin{equation}
D_i\equiv h^j_i\nabla_j
\end{equation}
is the covariant derivative projected onto the surface $S$ and
$h_{ij}\equiv g_{ij}-\xi_i\xi_j$ is the 3-dimensional geometric metric
tensor. From the above eqns, Ipser and Skivie$^{\cite{onze}}$ were
able to deduce the equation for the acceleration of a test particle
{\it off} the wall
\begin{equation}
\xi_iu^j\nabla_j\xi^i|_+=-\xi_iu^j\nabla_j\xi^i|_-=2\pi
G_N(\sigma^{eff}-2p^{eff}) 
\end{equation}
where $G_N$ is the Newtonian gravitational constant. By substitution
of $\sigma^{eff}$ and $p^{eff}_{||}$ into the eq. (26).
\begin{eqnarray}
\xi_iu^j\nabla_j\xi^i|_+ &=& 2\pi G_N(\sigma +\sigma_E-2\pi S^2-
2\sigma_E+4\pi S^2)\nonumber \\
&=& 2\pi G_N(\sigma -\sigma_E+2\pi S^2)
\end{eqnarray}
Since one knows that an observer who wishes to remain stationary next
to the wall must accelerate away from the wall if
$(\sigma^{eff}-2p^{eff}_{||})>0$ and towards to the wall if 
$(\sigma^{eff}-2p^{eff}_{||})<0$ one notes from (27) that repulsive
domain wall would have 
\begin{equation}
\sigma <\sigma_E-2\pi S^2
\end{equation}
Since in most of the physical systems in nature the spin energy
density $2\pi S^2$ is never higher than the surface electrostatic
energy density $\sigma_E$, $\sigma$ is not necessarily negative.
Nevertheless for attractive domain walls 
\begin{equation}
\sigma >\sigma_E-2\pi S^2>0
\end{equation}
and the stress-energy surface domain wall density should not violate
the weak energy condition, then the wormhole geometry is not
possible in this case.

Other applications of charged domain walls with spin and torsion maybe
found elsewhere. In particular a detailed account of the relation
between domain walls in EC-gravity and traversable wormholes maybe
found in recent paper by myself and Lemos$^{\cite{dez}}$.
Also recently other type of topological defect (cosmic strings) have
been investigated by Cl\'ement$^{\cite{vdois}}$ as a source of
flat wormholes. 

Finally it is important to mention that the matching conditions
in EC-theory were not 
necessary to deal with the non-Riemannian domain
walls$^{\cite{vtres}}$ discussed here. Distributional curvature of
cosmic strings has been  recently investigated by
Wilson$^{\cite{vquatro}}$. 

\section*{Acknowledgements}

\paragraph*{}
I am very much indebt to P.S. Letelier, Prof. C.A. Lopez and A. Wang
for helpful discussions on this subject of this paper. Some
suggestions of an unknown referee are gratefully acknowledged. Grants
from CNPq (Ministry of Science of Brazilian Government) and
Universidade do Estado do Rio de Janeiro (UERJ) are acknowledged.

\end{document}